\documentclass[twoside]{article}
\usepackage{epsfig.sty}
\newcommand{\beq}{\begin{eqnarray}}
\newcommand{\eeq}{\end{eqnarray}}
\begin{document}
\title{Sivers Differential Cross Section for $\pi+X$ Production via 
$p^{\uparrow}+p$ Collisions, and the Quark Sivers Function}
\author{Leonard S. Kisslinger\\
Department of Physics, Carnegie Mellon University, Pittsburgh, PA 15213\\
Ming X. Liu and Patrick McGaughey \\
P-25, Physics Division, Los Alamos National Laboratory, Los Alamos, NM 87545}
\date{}
\maketitle
\begin{abstract}
  We recently estimated pion+X production via proton collisions with a 
polarized proton target to enable the Collins fragmentation function to
be determined via future experiments. In the present article we first estimate
the Silvers differential cross section obtianed from experiments with the 
target proton having both positive and negative polarizations, and then
estimate the Sivers Function using quark distribution functions.
\end{abstract}
PACS Indices:12.38.Aw,13.60.Le,14.40.Lb,14.40Nd
\vspace{1mm}

\section{Introduction}

Recently we estimated the differential cross section for a proton colliding
with a polarized proton target to produce a pion and a hadron $X$, 
$\frac{d\sigma_{p^{\uparrow}p\rightarrow \pi X}}{dy}$, with $y$ the 
rapidity\cite{kmm16}, with a factor $D_{q^{\uparrow} \rightarrow q\bar{q}}$, the 
fragmentation probability for a polarized quark to fragment into a 
$q\bar{q}$\cite{bcfy95}. From this cross section,
and the fragmentation probability to produce a hadron from a transversely 
polarized quark, one can extract the Collins fragmentation 
function\cite{bggm08}.

  The present article is motivated in part by  the Letter of 
Intent by the E1039 collaboration\cite{LOI13-15}, with spokespersons from the 
Los Alamos National Laboratory, to measure the Sivers Function\cite{siv90}
via Drell-Yan experiments. A Drell-Yan process\cite{dy70} occurs in 
hadron-hadron scattering with a quark from one hadron and an antiquark from
the other hadron annihilating to create a boson.
A number of Deep Inelastic Scattering experiments\cite{hermes05,hermes09,
compass09,jlab11} have measured non-zero values for the Sivers Function, but
the E1039 Collaboration plan to make the first measurement of the
Sivers Function of sea quarks. 

  In the following section the Sivers Function obtained from  proton collision
experiments with a polarized proton target is discussed and the Sivers 
differential cross section is estimaterd. In the next section
the Sivers Function in terms of the quark distribution function in polarized 
protons is defined and used for our estimate of the Sivers Function.
\newpage
\section{The Sivers Function and the Sivers Cross Section}
  
The Sivers and Collins Functions in terms of experimental cross sections are
defined, e.g., in Refs\cite{hermes05,jlab11}. The Sivers function was defined
in terms of quark distribution functions with polarized protons in 
Ref.\cite{aam03}.

\subsection{Sivers and Collins Functions in terms of experiments}

For proton collisions with a polarized proton 
target producing a pion and a hadron (X), which is similar to the differential
cross section estimated in Ref\cite{kmm16} for the Collins Fragmentation
Function, except that both up and down proton polarizations are needed.
The Sivers and Collins Functions are given in terms of the assymmetry 
$A(\phi,\phi_s)$, see Refs.\cite{hermes05,jlab11}, where 
$\sigma_{p^{\uparrow}p\rightarrow \pi X},\sigma_{p^{\downarrow}p\rightarrow \pi X}$ are
cross sections for the production of $\pi$ + X via p-p scattering, with the
target proton polarization up, down,
\beq
\label{phi-phis}
  A(\phi_h,\phi_s)&=& \frac{1}{S_T}\frac{\sigma_{p^{\uparrow}p\rightarrow \pi X}
(\phi_h,\phi_s)-\sigma_{p^{\downarrow}p\rightarrow \pi X}(\phi_h,\phi_s)}
{\sigma_{p^{\uparrow}p\rightarrow \pi X}
(\phi_h,\phi_s)+\sigma_{p^{\downarrow}p\rightarrow \pi X}(\phi_h,\phi_s)}
\nonumber \\
    &\simeq& A_C sin(\phi_h+\phi_s)+A_S sin(\phi_h-\phi_s)  \; ,
\eeq 
where ($\phi_h,\phi_s$) are the azimuthal angles of the (hadron momentum, 
target spin) relative to the scattering plane, $S_T$ is the component of the 
target proton spin orthogonal to the scattering plane, $A_C$ is the Collins 
moment, and $A_S$ is the Sivers moment. Since the cross sections
$\sigma_{p^{\uparrow}p\rightarrow \pi X}(\phi_h,\phi_s),\sigma_{p^{\downarrow}p\rightarrow 
\pi X}(\phi_h,\phi_s)$ are very difficult for experiments to measure we
estimate the Sivers differential cross section in the next subsection section 
and the Sivers Function in the following section.

\subsection{Sivers Differential Cross Section}

From Eq(\ref{phi-phis}) we define the Sivers Differential Cross Section
$\frac{d\sigma^{Sivers}}{dy}$ as
\beq
\label{SivCrossSec}
 \frac{d\sigma^{Sivers}}{dy}&=& \frac{d\sigma_{p^{\uparrow}p\rightarrow \pi X}}
{dy}- \frac{d\sigma_{p^{\downarrow}p\rightarrow \pi X}}{dy} \; ,
\eeq
with 
From Ref\cite{kmm16} Eq(11) the differential cross section used to evaluate
the  Sivers differential cross section is, with $Aqq$ a known constant,
\beq
\label{ColCrossSec}
 \frac{d\sigma_{p^{\uparrow}p\rightarrow \pi X}}{dy}&=& Aqq [\Delta f_g(x(y),2m)
\Delta f_g(a/x(y),2m)+\Delta f_{d}(x(y),2m)\Delta f_{d}(a/x(y),2m) \nonumber \\
&&\Delta f_{\bar{d}}(x(y),2m)\Delta f_{\bar{d}}(a/x(y),2m)+
\Delta f_{u}(x(y),2m) \Delta f_{u}(a/x(y),2m) + \nonumber \\
&&\Delta f_{\bar{u}}(x(y),2m)\Delta f_{\bar{u}}(a/x(y),2m)]
 \frac{dx(y)}{dy} \frac{1}{x(y)} D_{q^{\uparrow} \rightarrow q\bar{q}} \; ,
\eeq
with rapidity $y$
\beq
\label{y(x)}
      y &=& \frac{1}{2} ln (\frac{E + p_z}{E-p_z}) \nonumber \\
        x(y) &=& 0.5 \left[\frac{m}{E}(\exp{y}-\exp{(-y)})+\sqrt{(\frac{m}{E}
(\exp{y}-\exp{(-y)}))^2 +4a}\right] \;  
\eeq
with $E,p_z$ the proton energy and momentum.
\newpage
From Ref.\cite{klm12} the distribution functions\cite{CTEQ6} for polarized
$p-p$ collisions for $Q=$ 3 GeV and E=200 GeV  
\beq
\label{dist-fcns}
  \Delta f_g(x)&=& 15.99-700.34 x+13885.4 x^2-97888. x^3 \nonumber \\
  \Delta f_{d}(x)&=& -5.378+205.6 x-4032.77 x^2 + 28371. x^3 \nonumber \\
 \Delta f_{u}(x)&=& 8.44-292.16 x+5675.16 x^2- 39722. x^3 \nonumber \\
  \Delta f_{\bar{u}}(x)&=& -1.447 +64.67 x-1268.24 x^2 +8878.32 x^3 \nonumber \\
  \Delta f_{\bar{d}}(x)&=&\Delta f_{\bar{u}}(x) \; ,
\eeq

with   $\Delta f_g$ the gluon distribution function and  $\Delta f_{q},
 \Delta f_{\bar{u}}$ the quark and anti-quark distribution functions.

\beq
\label{Dq}
   D_{q^{\uparrow} \rightarrow q\bar{q}}&=&\int_{0}^{1} dz
D_{q^{\uparrow} \rightarrow q\bar{q}}(z) \simeq 2.87 \times 10^{-3} \; , 
\eeq
with $D_{q^{\uparrow} \rightarrow q\bar{q}}(z)$ evaluated from Ref.\cite{bcfy95}

From Eqs(\ref{SivCrossSec},\ref{ColCrossSec}) the differential cross section 
$\frac{d\sigma^{Sivers}}{dy}$,  $D^{Sivers}=D_{q^{\uparrow} \rightarrow q\bar{q}}-
D_{q^{\downarrow} \rightarrow q\bar{q}}$, is
\beq
\label{SilCrossSec-final}
\frac{d\sigma^{Sivers}}{dy} &\propto& \frac{dx(y)}{dy} \frac{1}{x(y)}D^{Sivers} 
 \; .
\eeq

\beq
\label{DSivers}
   D^{Sivers}&=&\int_{0}^{1} dz
(D_{q^{\uparrow} \rightarrow q\bar{q}}(z)-D_{q^{\downarrow} \rightarrow 
q\bar{q}}(z)) \simeq 7.44 \times 10^{-3}\simeq 2.6\times D_{q^{\uparrow} 
\rightarrow q\bar{q}} \; , 
\eeq

Note also that for $-1 \leq y \leq 1$
\beq
\label{dist-ratio}
&&(1/\Delta f_g(x)\Delta f_g(a/x))\times [\Delta f_g(x)\Delta f_g(a/x)+
\Delta f_{d}(x)\Delta f_{d}(a/x)+\nonumber \\
&&\Delta f_{\bar{d}}(x)\Delta f_{\bar{d}}(a/x)+\Delta f_{u}(x) 
\Delta f_{u}(a/x(y),2m) + \nonumber\\
&&\Delta f_{\bar{u}}(x)\Delta f_{\bar{u}}(a/x))] \simeq 1.4
\eeq
Therefore, from Eqs(\ref{DSivers},\ref{dist-ratio})
\beq
\label{SilCrossSec-final-final}
\frac{d\sigma^{Sivers}}{dy} &\simeq&3.54\times \frac{d\sigma_{p^{\uparrow}p
\rightarrow \pi X}}{dy} \; .
\eeq

Note that in Ref\cite{kmm16} $\frac{d\sigma_{p^{\uparrow}p\rightarrow \pi X}}{dy}$
was calculated for E=200 Gev.

Therefore from Figure 2 in Ref\cite{kmm16} one finds $\frac{d\sigma^{Sivers}}
{dy}$
by simply multiplying the differential cross section in Figure 2 in
 Ref\cite{kmm16} by 3.54
\newpage

  $d\sigma^{Sivers}/dy$  for E=200 GeV is shown in figure 1
\vspace{2cm}

\begin{figure}[ht]
\begin{center}
\epsfig{file=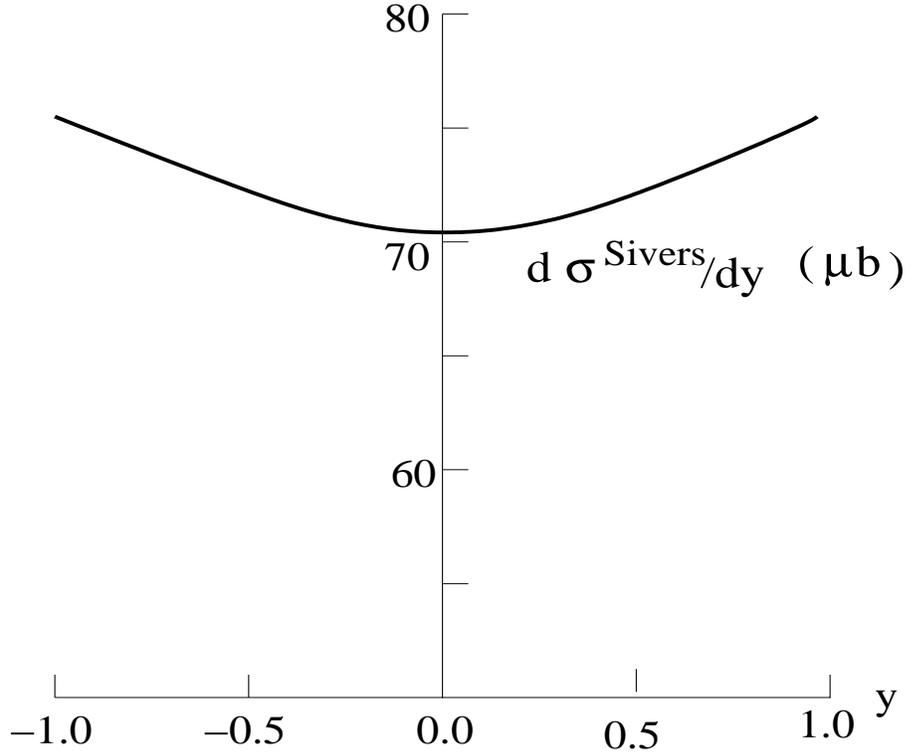,height=10cm,width=12cm}
\end{center}
\caption{ $d\sigma^{Sivers}/dy$ ($\mu$b) for E=200 GeV via $p^{\uparrow}p$ -
$p^{\downarrow}p$ collisions} 
\end{figure}

\section{Sivers Function in terms of Quark Distribution
  Functions}

The Sivers Function was defined in terms of quark distribution functions
in Refs\cite{aam03,badm04}. 

  The Sivers Function $f_S(z,k_T^2)$ is given by\cite{badm04}
\beq
\label{fsivers}
    f_{qp^{\uparrow}}(z,\vec{k}_T)-f_{qp^{\uparrow}}(z,-\vec{k}_T) &=&
 f_{qp^{\uparrow}}(z,\vec{k}_T)-f_{qp^{\downarrow}}(z,\vec{k}_T) \nonumber \\
  & =& -f_S(z,k_T^2) \frac{(\hat{P} \times \vec{k}_T\cdot \vec{S})}{2 M} \; ,
\eeq
where $\vec{P}$ is the proton momentum, $\vec{k}_T$ is quark momentum
perpendicular to $\vec{P}$,  $\vec{S}$ is the normalized proton spin,
with $S^2=-1.0$, and $M$ is the proton mass.
\newpage
  With the choice of frame the target rest frame, the Sivers Function can
be derived from the quark distribution function with polarized protons:
\beq
\label{fq} 
   f_S(z)&\simeq& (f_{qp^{\uparrow}}(z)-f_{qp^{\downarrow}}(z))(P/2M)
\nonumber \\
         &\simeq& (f_{qp^{\uparrow}}(z)-f_{qp^{\uparrow}}(-z))(P/2M) \; .
\eeq  

  Therefore, an approximate estimate of the Sivers Function can be obtained
from the quark fragmentation function $D_{q^{\uparrow} \rightarrow q\bar{q}}(z)$
from Ref\cite{kmm16}, which was obtained from Ref\cite{bcfy95} . 

\subsection{Estimate of the Sivers Function in terms of Quark 
Distribution Functions}

  We make an approximent estimate of the Sivers function $f_S(z)$ using
$f_{qp^{\uparrow}}(z) \simeq D_{q^{\uparrow} \rightarrow q\bar{q}}(z)$
   The estimate of $D_{q^{\uparrow} \rightarrow q\bar{q}}(z)$ from the Eq(31) in
Ref\cite{bcfy95} was discussed in Ref\cite{kmm16}.
\beq
\label{fqp-Dq}
  f_{qp^{\uparrow}}(z) &\simeq& D_{q^{\uparrow} \rightarrow q\bar{q}}(z) 
\simeq 2.125 \times 10^{-3} (6z+6z^2-15z^3-12z^4 +15z^5) \; ,
\eeq
with $0 \leq z \leq 1$. Therefore from Eqs(\ref{fq},\ref{fqp-Dq})
\beq
\label{fS(z)}
  f_S(z)&\simeq& 2.125 \times 10^{-3} \frac{P}{2M}(12 z -30 z^3 +30 z^5) \; .
\eeq
For E=$\sqrt{P^2+M^2}$=200 GeV and the proton mass M$\simeq$ 1 GeV, 
$P/2M\simeq 100$, giving
\beq
\label{fSfinal(z)}
  f_S(z)&\simeq& 0.2125(12 z -30 z^3 +30 z^5) \; .
\eeq 

The Sivers function, $f_S(z)$, is shown in Figure 2.
\vspace{1cm}
\begin{figure}[ht]
\begin{center}
\epsfig{file=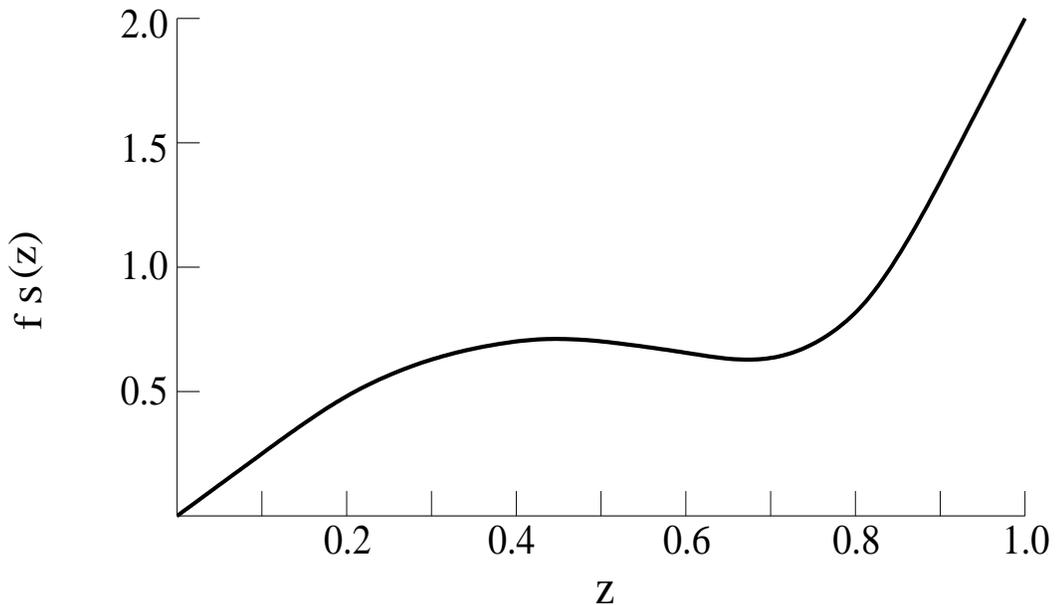,height=8cm,width=14cm}
\end{center}
\caption{$f_S(z)$=Sivers Function} 
\end{figure} 
\newpage

\section{Conclusions}

We have estimated the Sivers Differential Cross Section. As shown in
Figure 1 it should be possible to measure $\frac{d\sigma^{Sivers}}{dy}$ . 
We have also estimated the Sivers Function, shown in Figure 2.  

\vspace{1cm}

\Large{{\bf Acknowledgements}}\\
\normalsize
This work was supported in part by the DOE contracts W-7405-ENG-36 and 
DE-FG02-97ER41014. The author LSK was a visitor at Los Alamos National 
Laboratory, Group P25.  The authors declare that there is no conflict of 
interest regarding the publication of this paper.

\end{document}